\documentclass[a4paper,twoside]{article}
\newcommand{\hi}{{\sc H\,i}}

\newcommand{\affil}[1]{$^{\rm #1}$}
\usepackage[authoryear]{natbib}
\bibpunct{(}{)}{;}{a}{}{,}

\usepackage{graphicx}
\usepackage{grffile}
\date{} %Please leave the date blank
\title{\large\bf\flushleft Using Negative Detections to Estimate Source Finder Reliability}
\author{\parbox{\textwidth}{\flushleft
\vspace{-0.5cm}
{\it P. Serra\affil{A,D}, R. Jurek\affil{B} and L. Fl\"{o}er\affil{C}}\\
\vspace{0.4cm}
{\small \affil{A}\,Netherlands Institute for Radio Astronomy (ASTRON), Postbus 2, 7990 AA Dwingeloo, The Netherlands}\\
{\small \affil{B}\,CSIRO Astronomy \& Space Science, Australia Telescope National Facility, P.O. Box 76, Epping NSW 1710, Australia}\\
{\small \affil{C}\,Argelander-Institut f\"{u}r Astronomie, Auf dem H\"{u}gel 71, 53121 Bonn, Germany}\\
{\small \affil{D}\,Email: serra@astron.nl}}}
\begin{document}
\twocolumn[
\begin{changemargin}{.8cm}{.5cm}
\begin{minipage}{.9\textwidth}
\vspace{-1cm}
\maketitle
%
%
%%%%%%%%%%%%%     ABSTRACT    %%%%%%%%%%%%%
%Abstract of no more than 200 words here.
\small{\bf Abstract \rm

 We describe a simple method to determine the reliability of source finders based on the detection of sources with both positive and negative total flux. Under the assumption that the noise is symmetric and that real sources have positive total flux, negative detections can be used to assign to each positive detection a probability of being real.  We discuss this method in the context of upcoming, interferometric \hi\ surveys.}

%%%%%%%%%%%%%     KEYWORDS    %%%%%%%%%%%%%
\medskip{\bf Keywords:} methods: data analysis
% Please write all keywords in lower case. PASA uses the
% standard list of subject headings adopted by The Astrophysical Journal
% and available from http://www.journals.uchicago.edu/ApJ/keywords_text.html.
% Keywords are separated by em-dashes, i.e. ---

%%%%%%%%DO NOT EDIT%%%%%%%%%%%%
\medskip
\medskip
\end{minipage}
\end{changemargin}
]
\small
%%%%%%%%EDIT FROM HERE%%%%%%%%%%%%

\section{Introduction}
\label{intro}

\begin{figure*}[h]
\begin{center}
\includegraphics[width=16.5cm]{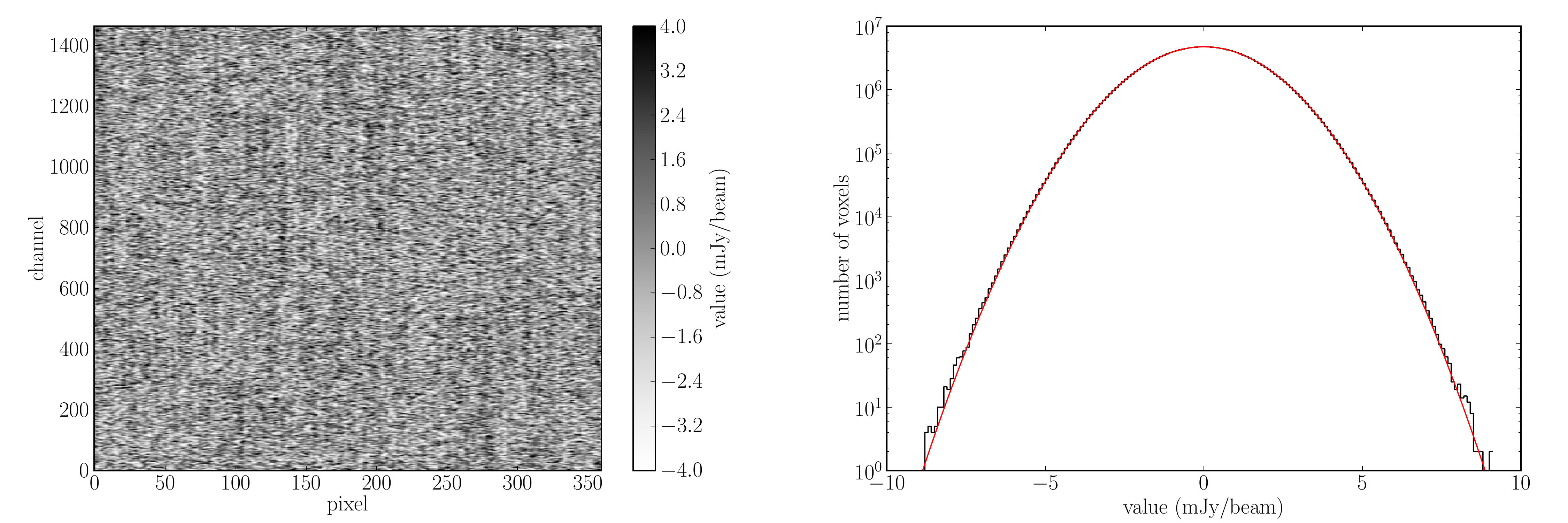}
\caption{Properties of the test noise cube. \it Left: \rm Right ascension-velocity slice going through the centre of the cube. Artefacts are visible as faint vertical stripes. \it Right: \rm Histogram of voxel values for the entire cube (black line) and a Gaussian distribution with $\sigma=1.6$ mJy beam$^{-1}$ (red line).}
\label{fig0}
\end{center}
\end{figure*}

\begin{figure*}[h]
\begin{center}
\includegraphics[width=16.5cm]{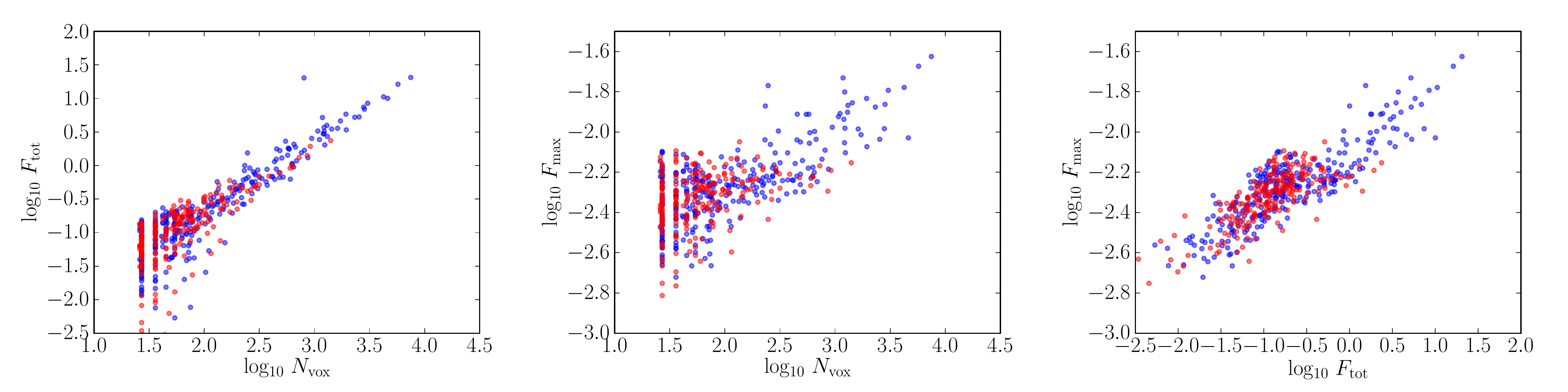}

\includegraphics[width=16.5cm]{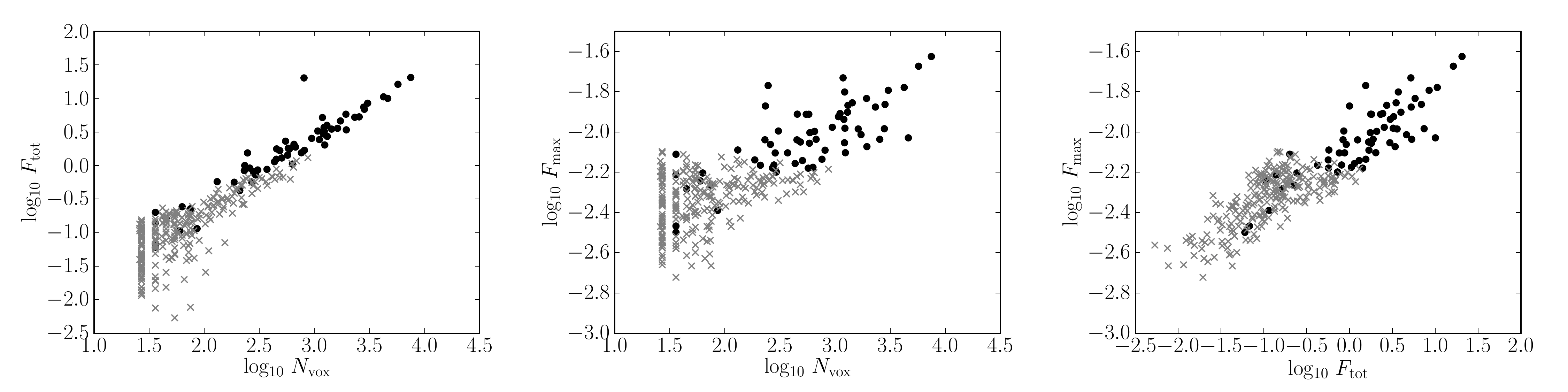}

\includegraphics[width=16.5cm]{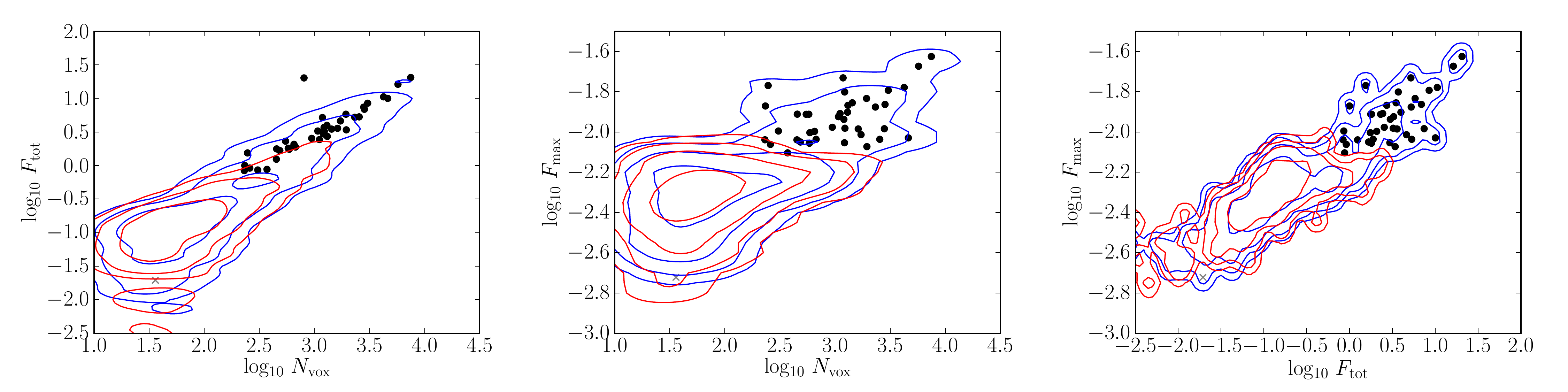}
\caption{Distribution of detections in the test \hi\ cube on all projections of the adopted parameter space (see text). \it Top\rm: Positive (blue) and negative (red) detections. \it Middle\rm: True (black circles) and false (grey crosses) positive detections. \it Bottom\rm: Same as middle panels, but showing detections with $R>0.99$ only. We also show constant surface-density contours of positive (blue) and negative (red) detections estimated from the distributions in the top panels as described in the text.}
\label{fig1}
\end{center}
\end{figure*}

\begin{figure*}[h]
\begin{center}
\includegraphics[width=16.5cm]{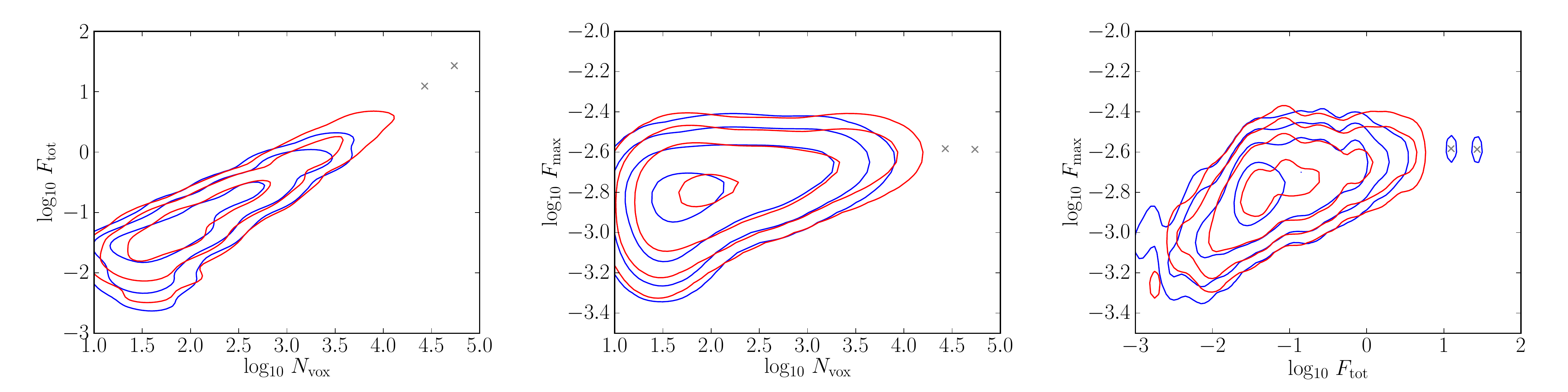}
\caption{Constant surface-density contours of positive (blue) and negative (red) detections for the datacube with RFI. Grey crosses indicate sources with $R>0.99$.}
\label{fig2}
\end{center}
\end{figure*}
In the coming years, a number of interferometric neutral-hydrogen (\hi) surveys will begin \citep[e.g.][]{2009pra..confE..14S,2009pra..confE..10V}. They will observe \hi\ within large cosmic volumes and detect tens of thousands of sources, many of which will be resolved both on the sky and in velocity.

These surveys will rely on automated source finders to detect objects present in the data. In particular, the detection of the faintest objects will require detection criteria close to the noise level. However, as fainter and fainter ``true'' objects are detected an increasing number of ``false'' detections will be included in the source catalogue (i.e., detections that are, in fact, noise peaks). Quantifying this effect is crucial to enable a proper scientific exploitation of the final \hi\ catalogues.

A quantity often used for this purpose is the reliability $R$ of a source catalogue. This is defined as:

\begin{equation}
R=\frac{T}{T+F},
\label{eq1}
\end{equation}

\noindent where $T$ and $F$ are the number of true and false detections, respectively. Normally, the price to pay for detecting faint objects is a decrease in $R$.

In some cases, a single value of $R$ may be used to characterise an entire source catalogue. However, it is more informative to study the $n$-dimensional function $R(p_1,p_2,...,p_n)$ where the $p_i$'s are a set of source parameters. For example, $R$ may be given as a function of objects' total flux and line width.

There are many ways of measuring $R$. \cite{2004MNRAS.350.1210Z} estimate the reliability of the HIPASS catalogue \citep{2004MNRAS.350.1195M} as a function of source total flux, peak flux and line-width by re-observing a sub-sample of the detected objects. They label confirmed detections as true and non-confirmed detection as false, and adopt a formalism equivalent to Eq. \ref{eq1} to estimate $R$. Unfortunately, this empirical procedure may not always be practical and it requires sources to be re-observed with at least the same data quality of the original observations.

Another technique is to create a dataset where model sources are injected on top of modelled (or observed) noise and run a source finder as one would do with the real data \citep[e.g.][]{2007ApJS..169..401K,2007AJ....133.2087S}. Detections corresponding to an input source are labelled as true and detections not corresponding to an input source are labelled as false. This approach gives a correct estimate of $R$ only if the model noise is a good approximation of the real noise and if model sources are representative of the objects actually contained in the data.

Here we discuss yet another method to measure $R$ based on the detection of ``negative'' sources, i.e., sources with negative total flux. This technique has been used in various forms by several authors working in different fields \citep[e.g.][]{2004ApJ...600L..99D,2004ApJ...612L..93Y,2009MNRAS.400..743K}. In this paper we develop it further with the aim of making it useful for future \hi\ surveys.

The main idea is to assume that true sources are ``positive'' (i.e., they have positive total flux) and that the noise is symmetric (we discuss the applicability of these assumptions in Section \ref{fut}). It follows that the number of false positive detections equals the number of negative detections. The reliability can then be defined as:

\begin{equation}
R=\frac{P-N}{P},
\label{eq2}
\end{equation}

\noindent where $P$ and $N$ are the number of positive and negative detections, respectively. It is trivial to verify that Eq. \ref{eq2} is equivalent to Eq. \ref{eq1} under the aforementioned assumptions.

The advantage of this method is that $R$ is measured directly from the data with no additional observational or modelling effort. In what follows we demonstrate this technique by applying it to a test \hi\ cube. We describe the cube and the source finder used for this purpose in Sections \ref{dataset} and \ref{sourcefinder}, respectively. In Section \ref{res} we illustrate the results. In Section \ref{fut} we discuss possible caveats and improvements of this technique. We draw conclusions in Section \ref{concl}.

\section{Test Data Cube}
\label{dataset}

We test the negative-source method on a data cube which is the sum of a noise cube and cubes containing only \hi\ sources. We build the noise cube by imaging in Stokes Q the continuum-subtracted visibility data obtained from a WSRT observation of the galaxy NGC~3941 \citep{2011arXiv1111.4241S}. \hi\ signal is unpolarised so the Stokes Q cube contains only noise (and imaging artefacts). We Fourier transform the visibilities using uniform weighting and 30-arcsec FWHM tapering. The resulting Gaussian beam has a FWHM of $\sim30\times30$ arcsec$^2$. The noise cube covers a sky area of 1 deg$^2$ and the recessional velocity range $\sim6000$ to $\sim12000$ km s$^{-1}$ ($z\sim0.02$--0.04; the median $z$ of galaxies detected by the WALLABY survey is expected to be $\sim0.03$, see \citealt{2009pra..confE..14S}). The channel width is $\sim3.8$ km s$^{-1}$, and we scale the cube to obtain a root-mean-square (r.m.s.) noise level of 1.6 mJy beam$^{-1}$, as per WALLABY specifications.

We add $\sim100$ \hi\ cubes available in the WHISP database \citep{2002ASPC..276...84V} to the noise cube. To do so we make use of the cubes' clean components derived as part of the WHISP data reduction. Each set of clean components representing an observed field is randomly redshifted within the $z$ range covered by the noise cube (using a triangular parent distribution for $z$), convolved with a $\sim30\times30$ arcsec$^2$ Gaussian beam, and placed at a random sky position within the noise cube (in a few cases this results in a position close to the edge of the cube). We note that some WHISP cubes contain more than one \hi\ source, so the number of input \hi\ sources is slightly larger than the number of WHISP cubes used.

This data cube is also used by \cite{jurek} to develop and refine the CNHI source finder and by \cite{westmeier} to test the Duchamp source finder. Compared to other test data cubes discussed in this issue \citep[e.g.][]{popping}, this cube has the advantage of including real interferometer noise (and therefore imaging artefacts) and real \hi\ sources. For example, the left panel in Figure \ref{fig0} shows a right ascension-velocity plane of the noise cube. Imaging artefacts are visible as vertical stripes on this projection. The right panel in the figure shows that the distribution of volumetric pixel (voxel) values is Gaussian.

\section{Source Finder}
\label{sourcefinder}

We look for objects in the test data cube by running a modified version of the \hi\ source-finder used in \cite{2011arXiv1111.4241S}. This finder smooths the data with a variety of kernels and, for each smoothed version of the cube,  detects signal above a specified threshold. In this way we attempt to optimise the signal-to-noise ratio of objects present in the data using a limited number of filters. In practice, we look for sources in the original \hi\ cube and in the cubes obtained by smoothing the original cube either on the sky, or in velocity, or along all three axes. In this study we use a Gaussian filter of FWHM=60 arcsec for smoothing on the sky, and a box filter of width 2, 4, 8, 16, and 32 channels for smoothing in velocity.

For each smoothed version of the cube we build a mask including all voxels brighter (in absolute value) than $4\sigma$, where $\sigma$ is the r.m.s. noise in that cube. The final mask is the sum of all masks (i.e., a voxel is included in the total mask if it is included in at least one of the individual masks). We size-filter the mask by performing morphological opening with the \it scipy.ndimage \rm\ Python package. We perform the opening using a $3\times3\times3$ structuring element where 1 pixel is 10 arcsec. Therefore, the structuring element has similar size as the beam and extends over 3 channels. Morphological opening filters out ensembles of voxels similar to or smaller than the structuring element. Therefore, this procedure removes most noise peaks included in the mask (noise peaks are typically smaller than the beam).

We merge detected voxels into individual sources using a $3\times3\times3$ structuring element. Because our detection criterion is applied to voxels' absolute value the final source catalogue includes both sources with positive and negative total flux.

The performance of this finder relative to other finders is discussed by \cite{popping}. They show that it detects more true sources than any other finder included in their study when applied to a test cube containing \hi\ disc model sources. Here we make use of the positive and negative source catalogue to estimate the reliability $R$ as a function of source parameters, and demonstrate how $R$ can be used to select samples of true detections.

\section{Results}
\label{res}

The top panels of Figure \ref{fig1} show the distribution of positive (blue) and negative (red) detections on three projections on the parameter space defined by source total flux $F_\mathrm{tot}$, peak flux $F_\mathrm{max}$, and number of voxels $N_\mathrm{vox}$\footnote{For negative detections $F_\mathrm{tot}$ and $F_\mathrm{max}$ are obtained after multiplying all voxels by $-1$. Both $F_\mathrm{tot}$ and $F_\mathrm{max}$ are given in Jy beam$^{-1}$.}. Positive detections are shown again in the middle panels of Figure \ref{fig1}, where black circles and grey crosses represent true and false detections, respectively. A detection is labeled true if its mask has non-zero overlap with an input source in the cube. Input sources are defined taking all voxels brighter than 0.16 mJy beam$^{-1}$ in the noise-less cube (1/10 of the noise level -- see Section \ref{dataset}) and merging them as in Section \ref{sourcefinder}. This results in 137 input sources.

We find 303 positive detections. Of these, 63 are true. We have verified that undetected input sources are too faint and occupy a different region of parameter space than detected ones. Figure \ref{fig1} shows that, \it a posteriori\rm, it would be easy to define a criterion to efficiently separate true from false detections for this particular combination of data cube and source finder. For example, all 41 detections with $\log_{10} F_\mathrm{tot}>-0.3$ and $\log_{10} F_\mathrm{max}>-2.1$ are true. Our goal is to show that a similar selection can be designed by applying Eq. \ref{eq2} to the distribution of positive and negative sources shown in Figure \ref{fig1}. The advantage of this second approach is that it needs no \it a priori\rm\ knowledge about the sources and can therefore be applied to any observed data cube.

We compute the density field of positive and negative detections by convolving their distribution shown in Figure \ref{fig1} with a Gaussian kernel of width $\sigma=0.075$, 0.035, 0.250 dex along the three logarithmic axes of the $\left(F_\mathrm{tot},F_\mathrm{max},N_\mathrm{vox}\right)$ space (we comment on the kernel choice below). We use the density fields to calculate the value of $P$ and $N$ at the location of each detected source, and apply Eq. \ref{eq2} to estimate the reliability $R$ at that location.

The bottom panels of Figure \ref{fig1} show the same distribution of points as in the middle panels, but for detections with $R>0.99$ only. Blue and red contours represent constant-surface-density contours of positive and negative sources, respectively. The figure shows that red and blue contours lay on top of each other in the noise-dominated region of the parameter space. Deviations occur in regions hosting true detections. We find 41 detections with $R>0.99$. Of these, 40 are true, in excellent agreement with the \it a posteriori\rm\ selection mentioned above. In fact, the only false detection (grey cross in the bottom panels) could be discarded based on its position in the parameter space.

We note that the choice of kernel made for the above calculation can influence the result of our analysis. A larger sample of detections would allow us to use a smaller kernel and, therefore, sample the function $R\left(F_\mathrm{tot},F_\mathrm{max},N_\mathrm{vox}\right)$ in a finer way. We attempt to make an objective choice of the kernel as follows.

We study the quantity $P-N$ estimated from positive and negative density fields at the location of negative detections. We assume that the noise dominates at these locations, so that the majority of detections are false. Given our assumptions (Section \ref{intro}), it then follows that $P=N$. Assuming that $P$ and $N$ follow a Poissonian distribution the quantity $(P-N)/\sqrt{P+N}$ should follow a Skellam distribution centred on zero and with variance 1 \citep{skellam}. However, for small kernels, the guaranteed presence of a negative source pushes the distribution to negative values. Only when the kernel is sufficiently large does the mean of the distribution move towards the expected value of zero. We therefore choose the smallest kernel which results in a $P-N$ distribution centred on zero.

\section{Caveats and Improvements}
\label{fut}

This method works under two basic assumptions: that true sources have positive flux and that the noise is symmetric (i.e., its distribution and morphology are symmetric about zero). The first assumption is not satisfied by data cubes where \hi\ absorption systems are also present. However, absorption is only detectable at the location of sufficiently bright continuum sources. Therefore, we believe that these systems could be easily excluded from an analysis like that presented in Section \ref{res}.

Deviations from noise symmetry may be a more serious issue. Real data can be thought as a superposition of \hi\ sources, perfect interferometer noise, and imaging artefacts resulting from faulty calibration, continuum subtraction, cleaning of bright sources and RFI removal. Such artefacts may represent a challenge for this method. The data cube analysed in Section \ref{res} contains real WSRT noise and includes some minor artefacts such as stripes visible in right ascension-velocity and declination-velocity projections (see Figure \ref{fig0}). However, it is a relatively clean case and does not allow us to assess the impact of imaging artefacts on the negative-source method.

To test the impact of RFI we analyse an \hi\ cube where RFI is present on short baselines. This cube is derived from a WSRT observation of NGC~3665 taken by \cite{2011arXiv1111.4241S}. Previous analysis has shown that no \hi\ emission is present in this cube. We run the same source finder described in Section \ref{sourcefinder} with the same settings, and perform the same analysis discussed in Section \ref{res}. The only difference is that we now use a Gaussian kernel of width $\sigma=0.10$, 0.10,0.25 dex along the three logarithmic axes of the $\left(F_\mathrm{tot},F_\mathrm{max},N_\mathrm{vox}\right)$ space. This is the smallest kernel for which the mean of all $P-N$ values at the location of negative sources equals zero (see Section \ref{res}).

The result is shown in Figure \ref{fig2}. Only 2 detections have $R>0.99$ (grey crosses). These are very large on the sky and their moment-0 image shows clearly that they are artefacts. We conclude that the method discussed here gives satisfactory results also in this particular case of RFI-contaminated data. The reason why our method may be able to deal with imaging artefacts is that they are usually symmetric in interferometric images, so that the method incorporates them as extra noise (in fact, the total flux of an interferometric dirty image is always zero because of the lack of data at zero spacing). A more thorough investigation of this aspect is beyond the scope of this paper and requires the analysis of a large number of data cubes including various types of imaging artefacts (e.g., RFI on different baselines and timescales, cleaning residual) in presence of true \hi\ sources.

This technique can be improved by working on a more appropriate parameter space. For example, we have characterised detected sources with the number of voxels they occupy. We could however consider more parameters describing the shape of a source. For example, the number of channels occupied by the source, and the major-to-minor axis ratio of the moment-0 image of the source. These parameters may be useful to separate spurious detections caused by imaging artefacts (which, for example, may be very elongated) from real sources. Analysis including more parameters will be possible with datasets larger than the one analysed here.

\section{Conclusions}
\label{concl}

We discuss a method to determine the reliability of sources detected in \hi\ cubes. We assume that true sources are positive and that the noise is symmetric. It follows that the number of negative detections equals the number of positive false detections. Negative detections can therefore be used to estimate the reliability $R$ of positive detections as a function of their position in a chosen source parameter space.

We demonstrate this method by running a smooth-and-clip source finder on a test \hi\ cube containing real interferometer noise and real \hi\ sources. We show that sources with $R>0.99$ are true. The volume of parameter space where this simple method gives $R>0.99$ is essentially the same which we would have selected knowing which source is true and which false in this test cube.

We discuss the applicability of this method to \hi\ cubes with artefacts. We show that at least in the analysed case of a cube with RFI the method performs well. The reason is that artefacts in interferometric images tend to be both positive and negative, so that they do not necessarily invalidate the noise symmetry assumption.

\section*{Acknowledgments}

The authors would like to thank Gyula J\'{o}zsa for useful discussion and Tom Oosterloo for a critical reading of the paper before submission.

\bibliographystyle{apj}
\bibliography{negsources}

\end{document}